\documentclass[12pt]{iopart}
\eqnobysec
\begin{document}
\title[Particle creation in the oscillatory phase of inflaton]
{  Particle creation in the oscillatory phase of inflaton}
\author{ P  K  Suresh }
\address 
{School of Physics, University of Hyderabad, Hyderabad-500 046.India.}
\ead{ pkssp@uohyd.ernet.in }
\begin{abstract}
A thermal squeezed state representation of inflaton is constructed for a flat Friedmann-Robertson-Walker background metric and the phenomenon of 
 particle creation  is examined during the oscillatory phase of inflaton, in the semiclassical theory of gravity. An approximate solution to the semiclassical Einstein 
equation is obtained in thermal squeezed state formalism by perturbatively and is found 
obey the same power-law expansion as that of classical Einstein equation. In addition to that the solution shows
oscillatory in nature except on a particular condition. 
It is also noted that, the 
coherently oscillating nonclassical inflaton, in thermal squeezed vacuum state, thermal squeezed state and thermal coherent state, suffer  
 particle production and the created particles exhibit oscillatory behavior. 
The present study can account for the post inflation particle creation due to thermal and quantum effects of inflaton in a flat FRW universe.
\end{abstract}
\section{Introduction}
According to the simplest version of the 
inflationary scenario, the universe in the past expanded  
exponentially with time, while its energy density was dominated  
by the effective potential energy density of a scalar field, 
called the inflaton. Sooner or later, inflation terminated and 
the inflaton field started quasiperiodic motion with slowly 
decreasing amplitude. The universe was empty of particles after 
inflation and particles of various kinds created due to the  
quasiperiodic evolution  of the inflaton field. The universe 
became hot again due the oscillations and decay of the created 
particles of various kinds. Form on,it can be described by the 
hot big bang theory.

        The  standard cosmology  provides reliable and 
tested account of the history of the universe from about 0.01sec 
after the big bang until today,  some 15 billions years later. 
Despite its success, the hot big bang model left many features of 
the universe unexplained. The most important of these are horizon 
problem, singularity problem, flatness problem, homogeneity 
problem, structure formation problem, monopole problem and so on. 
All these problems are very difficult and defy solution within 
the standard cosmology. Most of these problems have been either
 completely resolved or 
considerably relaxed in the context 
 inflationary scenario \cite{1}. At present 
there are  different versions \cite{2}-\cite{4} of the inflationary scenario. 
The main feature of all these versions is known as the 
inflationary paradigm. Inflationary cosmology is also widely accepted because of its success in explaining cosmological observations \cite{5}.

  Most of the inflationary scenarios are based on the classical gravity of the
Friedmann equation and the scalar field equation in the 
Friedmann-Robertson-Walker (FRW) universe, assuming  its validity even at the 
very early stage of the
universe. However, quantum effects of matter fields and quantum fluctuations 
are expected to play a significant role in this regime, though quantum gravity 
effects are still negligible. Therefore, the proper description of a 
cosmological
model can be studied in terms of the semiclassical gravity of the 
Friedmann equation with quantized matter fields as the source of gravity. 
The semiclassical quantum 
gravity  seems to be a viable method throughout the whole non-equilibrium 
quantum process from the pre-inflation period of hot plasma in thermal 
equilibrium to the inflation period and finally to the matter-dominated period.

Recently, the study of quantum properties of inflaton has been received 
much attention in semiclassical theory of gravity and inflationary scenarios.
 \cite{6,7}.  In the new 
inflation scenario \cite{8} quantum effects of the inflaton were partially taken into
account by using one-loop effective potential and an initial thermal 
condition.
In the stochastic inflation \cite{9} scenario the inflaton was studied quantum 
mechanically by dealing with the phase-space quantum distribution function and 
the probability distribution \cite{10}. 
 The aforementioned studies show that results obtained in classical gravity are 
quite different from those in semiclassical gravity. 
Such studies reveal that quantum effects  and quantum phenomena  
play an important role in inflation scenario and the related 
issues. Recently, it has been found that nonclassical state 
formalisms are quite useful to deal with quantum effects in 
cosmology \cite{11}-\cite{18}, particularly squeezed sates and coherent state formalism of quantum optics \cite{19}.   
  
 The above mentioned squeezed sates and coherent state formalisms are zero temperature states. There exist a thermal counterparts of coherent and squeezed sates and are useful to deal with finite  temperature effects and quantum effects.
  From the cosmological point of view it would be more natural to consider the temperature effects on the background of FRW metric. Therefore this motivates the study of  thermal squeezed states and thermal coherent states in cosmology.

 The goal of the  present paper is to study 
 quantum  and  finite temperature effects of  minimally coupled 
  massive inflaton in the FRW universe. Hence to examine 
  the  thermal and quantum
 particle creation, in the oscillatory phase, of the inflaton in 
thermal coherent and thermal squeezed state formalisms, in the semiclassical 
theory of gravity. 
For the present study we follow the unit system $ c= G= \hbar $=1.
\section{Thermal squeezed states and thermal coherent states} 
The thermo field dynamic \cite{20} formalism
 can be use to get the thermal counterparts of coherent and 
squeezed states.
The main feature of thermo field dynamics is the thermal Bogoliubov transformation that maps the theory from zero to finite temperature.
  One can construct a thermal vacuum 
$  \mid 0(\beta) \rangle $ annihilated by thermal annihilation operators and can express the average value of any observable  $A$ as the expectation value in the thermal \cite{20} vacuum
\begin{equation}
   Z(\beta)^{-1}\tr[\rho {A}]= \langle 0(\beta)\mid {A} \mid 0(\beta) \rangle ,
\end{equation}
where $\rho$ is the distribution function, $\beta= {1\over k T}$ and
$k$ Boltzmann's constant, and T the 
temperature. In order to fulfill the requirement (2.1), the vacuum should belong to the direct space between the original Fock space by an identical copy of it denoted by a tilde. Therefore 
\begin{eqnarray}
\mid 0(\beta) \rangle &=& e^{-iM} \mid 0, \tilde{0} \rangle,
M= -i\theta(\beta) ( a^{\dagger} \tilde{a}^{\dagger}-a\tilde{a} ),
\end{eqnarray}
where $a$, $a^{\dag}$ are the annihilation and creation  operators in original Fock space and $\tilde{a}$, $\tilde{a}^{\dag}$ are the same for the tilde space, and are obeying
boson commutation relations $[a,a^{\dag}]=[\tilde{a},\tilde{a}^{\dag}]=1$, the other combinations are
zero. 

The density matrix approach usually gives us a convenient method for 
incorporating finite temperature effects. Hence,
 various definitions of thermal coherent states (tcs) can be summarized by giving its density matrix and it can be written for the single mode case as \cite{21}
\begin{eqnarray}
\rho_{tcs} = D^{\dagger}(\alpha) e^{-\beta \omega 
a^{\dagger} a} D(\alpha),
\end{eqnarray}
where $\alpha$ is a complex number specifying the coherent state, $\omega $ is the energy of the mode, and  
\begin{eqnarray}
D(\alpha)&=& \exp{ \left(\alpha a^{\dagger}- \alpha^* a\right)}. 
\end{eqnarray}

The characteristic function for single mode 
thermal coherent state, ${\cal{Q}}_{tcs}$, is defined \cite{21} by 
\begin{eqnarray}
{\cal{Q}}_{tcs}(\eta ,\eta^*) &=& \exp [ - f(\beta) {\mid \eta \mid}^2
+ \eta^* \alpha - \eta \alpha^*],
\end{eqnarray}
where $\eta$ and $\eta^*$ are as independent variables and,
\begin{eqnarray}
f(\beta)& =& { 1 \over e^{\beta \omega} - 1}.
\end{eqnarray}

Similarly the density matrix for a single mode thermal squeezed states (tss) is given \cite{21}by
\begin{eqnarray}
\rho_{tss} = D^{\dagger}(\alpha) S^{\dagger} (\xi) e^{- \beta {a^{\dagger}} a
} S(\xi) D(\alpha),
\end{eqnarray}
where
\begin{eqnarray}
S(\xi) = \exp [( \xi {a^{\dagger} }^2 - \xi^* a^2 )/2],
 \xi = r e^{i \vartheta}.
\end{eqnarray}
Here $ r $ is the squeezing parameter  and $ \vartheta $ is the squeezing angle.

The characteristic function of a single mode thermal squeezed state is given by
\begin{eqnarray}
 \fl \eqalign{
{\cal{Q}}_{tss}(\eta ,\eta^*)=& \exp  [ -{\mid \eta \mid}^2
\left ( \sinh^{2} r \coth {\beta \omega \over 2} + f(\beta) \right ) \\ 
 & - {\cosh r \sinh r \over 2} \coth { \beta \omega \over 2} 
\left (
e^{-i\varphi} \eta^2 +e^{i \varphi} {\eta^*}^2 \right ) - \eta \alpha^*
+ \eta^* \alpha  ].}
\end{eqnarray}
 The  density matrix for a single mode thermal squeezed vacuum (tsv)is given by 
\begin{eqnarray}
\rho_{tsv}& =&  S^{\dagger} (\xi) e^{- \beta {a^{\dagger}} a
} S(\xi),
\end{eqnarray}
and the characteristic function is 
\begin{eqnarray}
\eqalign{
{\cal{Q}}_{tsv}(\eta ,\eta^*)=& \exp  [ -{\mid \eta \mid}^2
\left ( \sinh^{2} r \coth {\beta \omega \over 2} + f(\beta) \right ) \\ 
 & - {\cosh r \sinh r \over 2} \coth { \beta \omega \over 2} 
\left (
e^{-i\varphi} \eta^2 +e^{i \varphi} {\eta^*}^2 \right ) 
 ].}
\end{eqnarray}
Though the space is direct product between the original space and identical copy of it, the observational quantities are 
the expectation values of $ a, a^{\dagger}, a^2,  {a^{\dagger}}^2 $ \cite{21} etc.
Those quantities can be computed
in thermal coherent state, thermal squeezed state and thermal squeezed
vacuum state formalisms
by applying their corresponding characteristic function in the following relations.
\begin{eqnarray}
\eqalign{
\langle a \rangle =& {\partial {\cal{Q}} \over \partial 
{{\eta ^*}}}{ \mid_{ \eta
= {\eta^*} = 0}}, \\ 
\langle a^{\dagger } \rangle =& -{\partial {\cal{Q}} \over \partial
{\eta}}
{\mid_{ \eta
= \eta^*  = 0} }.}
\end{eqnarray}
Similarly the higher order expectation values of $ a $ and $a^{\dagger}$  can be also evaluated using the same procedure of eq (2.12). 
 \section{ Inflaton in a flat FRW metric}
 Consider a flat Friedmann-Robertson-Walker  spacetime with the line 
 element   
 \begin{equation}
 ds^{2}=-dt^{2}+R^{2}(t) (dx^{2}+dy^{2}+dz^{2}),
 \end{equation}
  The
metric is treated as an unquantized external
field.

 The minimally coupled inflaton with the gravity, for the metric (3.1), can be described by the Lagrangian 
\begin{equation} 
   L={1\over2} R^{3} \left({\dot{\varphi}^{2}}-m^{2}\varphi^{2}\right).
\end{equation}
Where overdot represents a derivative with respect to time.
 The equation governing the inflaton, for the metric (3.1), can be written as
 \begin{equation}
  \ddot\varphi+3\frac{\dot{R}}{R}\dot\varphi+m^{2}\varphi = 0.
\end{equation}
One can define the
 momentum conjugate to $\varphi$ as, 
$
\pi = \frac{\partial L}{\partial\dot{\varphi}}.
$
Thus, the Hamiltonian of the inflaton is 
\begin{equation}
H=\frac{\pi^{2}}{2R^{3}}+{1\over2}R^{3}m^{2}\varphi^{2}\,\,.
\end{equation}
Therefore, $0-0 $ component the energy-momentum tensor for the 
inflaton takes the following form
 \begin{equation} 
T_{00}=\frac{R^{3}}{2}(\dot\varphi^{2}+ m^{2}\varphi^{2})\,\,.
\end{equation}
 Consider the minimally coupled inflaton as the source of  gravity. Therefore the classical  Einstein 
equation  becomes 
\begin{equation}
\left(\frac{\dot{R}}{R}\right)^2=\frac{8\pi}{ 3} \frac{T_{00}}{R^3}
\,\,,
\end{equation}
where $T_{00}$ is the energy density of the inflaton, given by (3.5).
In  the cosmological context, the classical Einstein equation (3.6) means that 
the
Hubble constant, $H=\frac{\dot{R}}{R}$, is determined by the energy density of
the dynamically evolving inflaton as described by (3.3).  
 \section{ Thermal and quantum particle creation}
Since there is no consistent quantum theory of gravity available,
it would be meaningful to consider the semiclassical gravity theory
to study  quantum effect of matter field in a classical background metric. 
The
semiclassical  approach is also useful to deal with problems in 
cosmology, where quantum gravity effects are negligible. 
Oscillatory phase of inflaton is such a situation, where one can neglect the quantum gravity effects. 
 Therefore the present 
study  can be restricted 
in the frame work of semiclassical theory of  gravity.
 In semiclassical theory the Einstein equation can be written as 
  \begin{equation}
G_{\mu\nu}=8\pi \langle {\hat{T}}_{\mu\nu}\rangle\,\,,
\end{equation}
where the quantum field, represented by a scalar field  $\phi$, is 
governed by the time-dependent Schr$\ddot{o}$dinger equation
\begin{equation}
i\frac{\partial \phi}{\partial t} =\hat{H}_{\phi} \phi
\,\,.
\end{equation}.
Consider
 quantum  inflaton as the source, then the Friedmann equation, for the metric (3.1), 
in the semiclassical theory, can be written as
\begin{equation}
\left(\frac{\dot{R}}{R}\right)^2=\frac{8\pi}{3} \frac{1}{R^3}
\langle \hat{H}_\varphi \rangle\,\,,
\end{equation}
where $\langle \hat{H}_{\varphi} \rangle$ represent the expectation value of the Hamiltonian
of the inflaton in a quantum state under consideration.

 The  inflaton  can be
described by the time dependent harmonic oscillator, with the 
Hamiltonian given in (3.4).
To study,  the semiclassical Friedmann equation, the 
expectation value 
the  Hamiltonian (3.4) to be computed, in a quantum state under consideration. Therefore (3.4) becomes
\begin{equation}
\langle\hat{H}_{\varphi} \rangle=\frac{1}{2R^{3}}\langle\hat{\pi}^2\rangle+ 
\frac{m^{2}R^{3}}{2}\langle\hat{\varphi}^2\rangle\,\, .
\end{equation}
The eigenstates of the  Hamiltonian are the Fock states
\begin{equation}
a^{\dag}(t)a(t)|n,\varphi,t \rangle = n|n,\varphi,t \rangle\,\,,
\end{equation}
where
\begin{eqnarray}
\eqalign{
a(t)=&\varphi^*(t) \hat{\pi} - R^{3}\dot{\varphi}^*(t) 
\hat{\varphi},\\
a^{\dag}(t)=&\varphi(t) \hat{\pi}-R^{3} \dot{\varphi}(t)
\hat{\varphi}\,\,. }
\end{eqnarray}
As an  alternative to the $n$  representation, consider
the inflaton  in  thermal squeezed state formalism. Therefor the expectation value of the Hamiltonian (4.4) in thermal squeezed  state can be computed as follow.

 From  (2.9), (2.12) and (4.6), we get  
 \begin{equation}
 \eqalign{
  \langle  \hat{\pi}^{2}\rangle =&
  - R^{6}\dot{\varphi}^{2}\left( \alpha^{2}-e ^{i\vartheta}\cosh r \sinh r \coth \frac{\beta \omega}{2} \right) \\ 
&-R^{6}\dot{\varphi}^{\ast 2} \left( \alpha^{\ast 2}-e ^{-i\vartheta}\cosh r \sinh r \coth \frac{\beta \omega}{2}\right)\\ 
&+ R^{6} \dot{\varphi} ^{\ast}\dot{\varphi} \left( 2| \alpha|^{2}+ 2 \sinh^{2} r \coth\frac{\beta \omega}{2} + 2 f(\beta) + 1 \right),}
\end{equation}  
and
\begin{eqnarray}
 \eqalign{
  \langle  \hat{\varphi}^{2}\rangle=&
  - \varphi^{2}\left( \alpha^{2}-e ^{i\vartheta}\cosh r \sinh r \coth \frac{\beta \omega}{2} \right) \\ 
&-\varphi^{\ast 2} \left( \alpha^{\ast 2}-e ^{-i\vartheta}\cosh r \sinh r \coth \frac{\beta \omega}{2}\right)\\ 
&+  \varphi ^{\ast}\varphi \left( 2| \alpha|^{2}+ 2 \sinh^{2} r \coth\frac{\beta \omega}{2} + 2 f(\beta) + 1 \right).}
\end{eqnarray}  
 Substituting (4.7) and (4.8) in (4.4), and the apply the result in (4.3), then the
semiclassical Friedmann equation becomes
\begin{eqnarray}
 \fl \eqalign{
  \left(\frac{\dot{R}}{R}\right)^2 =&
  \frac{4 \pi}{3} \left[ (\dot{\varphi }^{\ast} \dot{\varphi} + m^2 {\varphi }^{\ast} \varphi)
 \left( 2| \alpha|^{2}+ 2 \sinh^{2} r \coth\frac{\beta \omega}{2} + 2 f(\beta) + 1 \right) \right. \\ 
&\left.- (\dot{ \varphi }^{2} +m^{2} \varphi^{2})
\left( \alpha^{2}-e ^{i\vartheta}\cosh r \sinh r \coth \frac{\beta \omega}{2} \right)  \right.\\ 
&\left.-(\dot{\varphi}^{\ast 2}+ m^2 \varphi^{\ast 2})
\left( \alpha^{\ast 2}-e ^{-i\vartheta}\cosh r \sinh r \coth \frac{\beta \omega}{2}\right) \right],}
\end{eqnarray} 
where, $\varphi$ and $\varphi^*$ satisfy  eq (3.3) and the 
Wronskian condition
\begin{equation}
R^3(t)\left(\dot{\varphi^*}(t)\varphi(t)-\varphi^{*}(t) \dot{\varphi}(t) \right)=i\,\,.
\end{equation}
The above boundary condition, fixes the normalization constants 
of the  two independent solutions.

To solve the self-consistent semiclassical Einstein equation (4.9),
transform the solution
in the following form
\begin{equation}
\varphi(t)=\frac{1}{R^{3\over2}}\psi(t),
\end{equation}
thereby obtaining
\begin{equation}
\ddot{\psi}(t)+\left(m^{2}-{3\over4}\left(\frac{\dot{R}(t)}{R(t)}\right)^{2}
-{3\over2}
\frac{\ddot{R}(t)}{R(t)}\right)\psi(t)=0\,\,.
\end{equation}
Next, focus on the oscillatory phase of the inflaton after inflation. In the
parameter region satisfying  the inequality
\begin{equation}
m^2 > \frac{3\dot{R}^2}{4R^2}+\frac{3\ddot{R}}{2R},
\end{equation}
the inflaton has an oscillatory solution of the form
\begin{equation} 
\psi(t)=\frac{1}{\sqrt{2\sigma(t)}}\exp(-i\int \sigma(t)dt)\,\,.
\end{equation}
With
\begin{equation}
\sigma(t)=\sqrt{m^{2}-{3\over4}\left(\frac{\dot{R}}{R}\right)^{2}-{3\over2}
\frac{\ddot{R}}{R}+{3\over4}\left(\frac{\dot{\sigma}(t)}{\sigma(t)}\right)^{2}-
{1\over2}\frac{\ddot{\sigma}(t)}{\sigma(t)}}\,\,.
\end{equation}
By applying the transform solution (4.11) in (4.9), and also using the fact
$ \alpha = e^{i\delta} \alpha $,
 we obtain
\begin{eqnarray}
\fl \eqalign{
 R(t)=&\left[\frac{2 \pi}{3 \sigma} \frac{1} {(\frac{\dot{R}}{R})^2} \left[\frac{1}{4}\left(\left( 3 \frac{\dot{R}}{R} + \frac{\dot{\sigma}}{\sigma}\right)^{2}
+\sigma^{2} +m^{2}\right)\right. \right.\\
&\left. \left. \times
\left( 2 |\alpha |^{2} + 2 \sinh^2 r \coth \frac{\beta \omega}{2}
+ 2 f(\beta) + 1 \right)\right. \right.\\
&\left. \left.
-\frac{1}{4}\left(\left( 3 \frac{\dot{R}}{R} + \frac{\dot{\sigma}}{\sigma}\right)^{2}
-\sigma^{2} +m^{2}\right) \right.\right.\\
&\left.\left. \times \left( 2 \alpha^{2} \cos(2 \delta - 2 \sigma t ) 
- \cos(\vartheta - 2 \sigma t)
\sinh(2r) \coth \frac{\beta \omega}{2}\right) \right. \right.\\
& \left. \left.+ \sigma \left( 3 \frac{\dot{R}}{R} + \frac{\dot{\sigma}}{\sigma}\right)
 \left( 2 \alpha^{2} \sin(2 \delta - 2 \sigma t ) 
+ \sin(\vartheta - 2 \sigma t)
\sinh(2r) \coth \frac{\beta \omega}{2}\right)\right]
\right ]^{1/3}.}
\end{eqnarray}
The next order approximation solution of the eq (4.16) can be obtained by using the  following approximation 
ansatzs 
\begin{equation}
\sigma_{0}(t)=m,  
\end{equation}
and
\begin{equation}
R_{0}(t)=R_{0}t^{2\over3}\,\,.
\end{equation}
Thus  we get
\begin{eqnarray}
 \fl \eqalign{
 R_{1}(t)=&\left[3 \pi m t^{2} \left[ 
 \left( 1+ \frac{1}{2 m^2 t^2}\right)
\left( 2 |\alpha |^{2} + 2 \sinh^2 r \coth \frac{\beta \omega}{2}
+ 2 f(\beta) + 1 \right)\right. \right.\\
&\left. \left.
-\frac{1}{2 m^2 t^2} \left( 2 \alpha^{2} \cos(2 \delta - 2 m t ) 
- \cos(\vartheta - 2 m t)
\sinh(2r) \coth \frac{\beta \omega}{2}\right) \right. \right.\\
& \left. \left.+ \frac{2}{m t^2}
 \left( 2 \alpha^{2} \sin(2 \delta - 2 m t ) 
+ \sin(\vartheta - 2 m t)
\sinh(2r) \coth \frac{\beta \omega}{2}\right)\right]
\right ]^{1/3}.}
\end{eqnarray}
When $2 \delta = \vartheta= 2 m t$, then (4.19) becomes
\begin{eqnarray}
 \fl \eqalign{
 R_{1}(t)=&\left[3 \pi m t^{2} \left[ 
 \left( 1+ \frac{1}{2 m^2 t^2}\right)
\left( 2 |\alpha |^{2} + 2 \sinh^2 r \coth \frac{\beta \omega}{2}
+ 2 f(\beta) + 1 \right)\right. \right.\\
&\left. \left.
-\frac{1}{2 m^2 t^2} \left( 2 \alpha^{2} 
- 
\sinh(2r) \coth \frac{\beta \omega}{2}\right) \right]
\right ]^{1/3}.}
\end{eqnarray}
Next, consider the particle production of the inflaton, in 
thermal squeezed states formalisms, in semiclassical theory 
of gravity. First,  consider the Fock space which has a one 
parameter dependence on the cosmological time $t$. The number of 
particles at a later time $t$ produced from the vacuum at the 
initial time $t_{0}$ is given by
\begin{equation}
N_0(t,t_0)=\langle 0,\varphi,t_0\mid\hat{N}(t)\mid 0,\varphi, t_0\rangle ,
\end{equation}
here, $\hat{N}(t)=a^{\dag}a$ and its expectation value   and can
be calculated by using (4.6). Therefor,
\begin{equation}
\langle \hat{N}(t) \rangle = R^6 \dot{\varphi} \dot{\varphi^*} \langle \hat{\varphi^2}  \rangle
+ \varphi \varphi^* \langle \hat{\pi}^2   \rangle - R^3 \varphi 
\dot{\varphi}^* \langle \hat{\pi} \hat{\varphi} \rangle-R^3 \dot{\varphi} \varphi^* \langle
\hat{\varphi} \hat{\pi}
\rangle .
\end{equation} 
Again using (4.6) we get
\begin{equation}
 \eqalign{
 \langle\hat{\varphi}^2\rangle&=\varphi^* \varphi ,   \\
\langle\hat{\pi}^2\rangle&=R^6\dot{\varphi}^*\dot{\varphi}, \\
\langle\hat{\pi}\hat{\varphi}\rangle&=R^3 \dot{\varphi} \varphi^*, \\ 
\langle\hat{\varphi}\hat{\pi}\rangle&=R^3\varphi\dot{\varphi}^* . } 
\end{equation}
  Therefore, substituting (4.23), in (4.22), we get
\begin{equation}
N_0(t,t_0)=R^{6}|\varphi(t)\dot{\varphi}(t_0)-\dot{\varphi}(t)\varphi(t_0)|^{2}\,\,.
\end{equation}
 Using the  approximation 
ansatzs (4.17),(4.18)
and (4.24),
the number of particles
created at a later time $t$  from the vacuum state at the initial time
 $t_0$ in the limit $mt_0$,  $mt>1$ can be computed and is [7] given by
\begin{eqnarray}
 \nonumber  N_0(t,t_0)&=&
 \frac{1}{4 \sigma(t) \sigma(t_0)}\left(\frac{R(t)}{R(t_0)}\right)^3  
\left[\frac{1}{4}\left(3\frac{\dot{R}(t)}{R(t)}-3\frac{\dot{R}(t_0)}{R(t_0)} \right.\right.\\
\nonumber && \left.\left.
-\frac{\dot{\sigma}(t)}{\sigma(t)}+\frac{\dot{\sigma}(t_0)}{\sigma(t_0)}\right)^2
+ (\sigma(t)-\sigma(t_0))^2\right] \\
 &\simeq &\frac{ {( t-t_{0})}^2 } { 4 m^2 t_{0}^4} \, .
\end{eqnarray}
To compute the particle creation in thermal squeezed state, the expectation values of the 
$\langle\hat{\pi}^2\rangle , \langle\hat{\varphi}^2\rangle, \langle\hat{\pi} \hat{\varphi}  \rangle  $ and $\langle \hat{\varphi} \hat{\pi}\rangle$   in the thermal squeezed state are required. And are respectively  obtained by using eqs (2.9),(2.12) and  (4.6) as follow
 \begin{eqnarray} 
\nonumber \fl \langle\hat{\pi}^2\rangle_{tss} = - R^{6} \dot{\varphi}^{2}(t_{0})\left( \alpha^{2}-e ^{i\vartheta}\cosh r \sinh r \coth \frac{\beta \omega}{2} \right) \\
\nonumber -R^{6}\dot{\varphi}^{\ast 2}(t_{0}) \left( \alpha^{\ast 2}-e ^{-i\vartheta}\cosh r \sinh r \coth \frac{\beta \omega}{2}\right) \\
\nonumber + R^{6} \dot{\varphi} ^{\ast}(t_{0})\dot{\varphi} (t_{0})\left( 2| \alpha|^{2}+ 2 \sinh^{2} r \coth\frac{\beta \omega}{2} + 2 f(\beta) + 1 \right) 
,   \\
\nonumber \fl \langle\hat{\varphi}^2\rangle_{tss}= - \varphi^{2}(t_{0})\left( \alpha^{2}-e ^{i\vartheta}\cosh r \sinh r \coth \frac{\beta \omega}{2} \right) \\
\nonumber -\varphi^{\ast 2} (t_{0})\left( \alpha^{\ast 2}-e ^{-i\vartheta}\cosh r \sinh r \coth \frac{\beta \omega}{2}\right)\\ 
\nonumber +  \varphi ^{\ast}(t_{0})\varphi (t_{0})\left( 2| \alpha|^{2}+ 2 \sinh^{2} r \coth\frac{\beta \omega}{2} + 2 f(\beta) + 1 \right), \\ 
\eqalign{
\fl \langle\hat{\pi}\hat{\varphi}\rangle_{tss}=  -R^3
\varphi(t_0)\dot{\varphi}(t_0)
\left( \alpha^{2}-e ^{i\vartheta}\cosh r \sinh r \coth \frac{\beta \omega}{2} \right)\\
-R^3\varphi^*(t_0)
 \dot{\varphi}^*(t_0) \left( \alpha^{\ast 2}-e ^{-i\vartheta}\cosh r \sinh r \coth \frac{\beta \omega}{2}\right)   \\ 
  +R^3  (
 \varphi(t_0)\dot{\varphi}^*(t_0)  
+  \varphi^*(t_0)\dot{\varphi}(t_0) )
\left( 2| \alpha|^{2}+ 2 \sinh^{2} r \coth\frac{\beta \omega}{2} + 2 f(\beta) + 1 \right) 
,\\  
\fl \langle\hat{\varphi}\hat{\pi}\rangle_{tss} =  -R^3
\dot{\varphi}(t_0)\varphi(t_0)
\left( \alpha^{2}-e ^{i\vartheta}\cosh r \sinh r \coth \frac{\beta \omega}{2} \right)\\
-R^3
 \dot{\varphi}^*(t_0)\varphi^*(t_0) \left( \alpha^{\ast 2}-e ^{-i\vartheta}\cosh r \sinh r \coth \frac{\beta \omega}{2}\right)   \\ 
  +R^3  (\dot{\varphi}^*(t_0) 
 \varphi(t_0) 
+ \dot{\varphi}(t_0) \varphi^*(t_0) )
\left( 2| \alpha|^{2}+ 2 \sinh^{2} r \coth\frac{\beta \omega}{2} + 2 f(\beta) + 1 \right).  }
\end{eqnarray}
Substituting (4.26) in (4.22), we get
\begin{eqnarray}
\eqalign{
 N_{tss}(t,t_{0})=& \frac{1}{16}\frac{1}{\sigma(t)}\frac{1}{\sigma (t_0)}\left(\frac{R(t)}{R(t_0)}\right)^3 \\ 
& \times \left [ \left [\left( 3 \frac{\dot{R}(t)}{R(t)}
- 3 \frac{\dot{R}(t_0)}{R(t_0)}  
 + \frac{\dot{\sigma}(t)}{\sigma(t)}
- \frac{\dot{\sigma}(t_0)}{\sigma(t_0)}\right)^2  + \sigma (t)^{2}-\sigma(t_0)^{2}\right]  \right. \\ 
& \left. 
 \times \left( 2| \alpha|^{2}+ 2 \sinh^{2} r \coth\frac{\beta \omega}{2} + 2 f(\beta) + 1 \right) \right.\\ 
& \left.  -\left( 3 \frac{\dot{R}(t)}{R(t)}
- 3 \frac{\dot{R}(t_0)}{R(t_0)} + \frac{\dot{\sigma}(t)}{\sigma(t)}
- \frac{\dot{\sigma}(t_0)}{\sigma(t_0)}\right)^2 \right.\\ 
&\left. \times \left[
\left( \alpha^{2}-e ^{i\vartheta}\cosh r \sinh r \coth \frac{\beta \omega}{2} \right) e^{-2 i \sigma(t_0) t_0}\right. \right.\\ 
&\left.\left.  + \left( \alpha^{\ast2}-e ^{-i\vartheta}\cosh r \sinh r \coth \frac{\beta \omega}{2} \right) e^{2 i \sigma(t_0) t_0}\right] 
\right ].}
\end{eqnarray} 
Which is 
 the number of particles produced in thermal  squeezed  
state,  at a later time $t$   from 
 the initial time
$t_0$.

By using (4.17) and (4.18) the above equation (4.27) can be rewritten as follows
\begin{eqnarray}
\eqalign{
 N_{tss} \simeq &\frac{1}{4}\frac{1}{m^{2}}\frac{(t-t_0)^2}{t_0 ^4} \left[
\left( 2| \alpha|^{2}+ 2 \sinh^{2} r \coth\frac{\beta \omega}{2} + 2 f(\beta) + 1 \right) \right.\\ 
&\left.-
 \left( \alpha^{2}-e ^{i\vartheta}\cosh r \sinh r \coth \frac{\beta \omega}{2} \right) e^{-2 i m t_0} \right.\\ 
&\left.- \left( \alpha^{\ast2}-e ^{-i\vartheta}\cosh r \sinh r \coth \frac{\beta \omega}{2} \right) e^{2 i m t_0}\right].}
\end{eqnarray}
By taking $ \alpha = e^{i\delta} \alpha $, eq (4.28) becomes 
\begin{eqnarray}
\eqalign{
 N_{tss} \simeq& N_0(t,t_0)
\left[
 2|\alpha|^{2}+ 2 \sinh^{2} r \coth\frac{\beta \omega}{2} + 2 f(\beta) +1  \right.\\
&\left. -2 \alpha^{2} \cos (2\delta-2  m t_0)+  \cos (\vartheta -2  m t_0) \sinh (2r) \coth \frac{\beta \omega}{2}   \right],}
\end{eqnarray}
where $N_0(t,t_0)$ is given by (4.25).

When $\alpha = 0$ eq (4.29) leads to 
\begin{eqnarray}
\eqalign{
 N_{tcs} \simeq& N_0(t,t_0)
\left[
 2|\alpha|^{2} + 2 f(\beta) +1  
 -2 \alpha^{2} \cos (2\delta-2  m t_0)  \right]}.
 \end{eqnarray}
 Which is particle creation in thermal coherent state. The same result can be also obtained by using eqs (2.5), (2.12), (4.6), (4.17), (4.18) and (4.22).

When $r=0$,  eq (4.29) becomes 
\begin{eqnarray}
\eqalign{
 N_{tsv} \simeq& N_0(t,t_0)
\left[
  2 \sinh^{2} r \coth\frac{\beta \omega}{2} + 2 f(\beta) +1  \right.\\
 &\left.+  \cos (\vartheta -2  m t_0) \sinh (2r) \coth \frac{\beta \omega}{2}   \right].}
\end{eqnarray}
The eq (4.31) can be also obtained by using eqs (2.11), (2.12), (4.6), (4.17), (4.18) and  (4.22), and is the particle production due thermal squeezed vacuum state.

 When $2\delta=2  m t_0$ and $\vartheta =2  m t_0$,
eqs (4.29), (4.30) and (4.31) respectively become
\begin{eqnarray}
\eqalign{ N_{tss} \simeq &N_0(t,t_0)
\left[
 2| \alpha|^{2}+ 2 \sinh^{2} r \coth\frac{\beta \omega}{2} + 2 f(\beta) +1  \right.\\
&\left.   -2
  \alpha^{2} + \sinh (2r) \coth \frac{\beta \omega}{2} 
\right],}
\end{eqnarray}
\begin{eqnarray}
 N_{tcs} \simeq &N_0(t,t_0)
\left[
 2| \alpha|^{2}+  2 f(\beta) +1  
-2 \alpha^{2}
  \right],
\end{eqnarray}
and
\begin{eqnarray}
 N_{tsv} \simeq N_0(t,t_0)
\left[
  2 \sinh^{2} r  + 2 f(\beta) +1+
  \sinh (2r) \coth\frac{\beta \omega}{2}     
\right].
\end{eqnarray}
When $r= \alpha =0$, then eq (4.29) take the following form
\begin{eqnarray}
 N_{th} \simeq &N_0(t,t_0)
\left[
  2 f(\beta) +1  
  \right].
\end{eqnarray}
Which is the particle creation due to purely thermal effects. 
{\section{ Conclusions} }
 In this paper, we studied  particle production
of the coherently oscillating inflaton, after the inflation, in thermal coherent states
  and thermal squeezed states formalisms, in the frame work of semiclassical theory
of gravity. 
 The number of 
particles at a
later time $t$,  produced from the thermal coherent state, at the initial time $t_0$, 
in the limit $mt_0$, $mt > 1$   calculated. 
It shows, the particle production depends on the 
coherent  state parameter and finite temperature effects. The particle creation in thermal squeezed vacuum state in the 
limit   $mt_0 >mt >1$ is also computed, it is found that the 
particle production depending on the associated squeezing parameter and temperature. Similarly
the number of particles produced in thermal squeezed state  also  computed.
 It is observed
that, when $r=0$, the result agree with the number of particles produced in the
thermal coherent state
 and when $\alpha = 0$, the result equal to the number of particles
created in thermal  squeezed vacuum state.

 The approximate leading solution obtained for the Einstein equation, in the thermal squeezed sates  shows oscillatory behavior except when the condition, $ 2\delta = \vartheta = 2 mt$, satisfies. 
Though both 
classical and quantum inflaton in the oscillatory phase of the 
inflaton lead  the same power law expansion, the correction to 
the expansion does not show any oscillatory behavior in 
semiclassical gravity in contrast to the oscillatory behavior 
seen in classical gravity only when $ 2\delta = \vartheta = 2 mt$. It is also noted that, the 
coherently oscillating  inflaton, in thermal squeezed vacuum, thermal squeezed and thermal coherent states representation, suffer  particle creation  and  created particle exhibit oscillations.  The oscillation of the created particles is necessary to preheat the universe to hot again after the inflation. The present study can account for the post inflation particle creation due to thermal and quantum effects of inflaton in a flat FRW universe. Since the created particle oscillate, we hope that this kind of study can light on  preheating issues of post inflationary scenario.

\end{document}